\documentclass[prb,twocolumn,preprintnumbers,amsmath,amssymb,showpacs,nofootinbib,floatfix]{revtex4}

\usepackage{graphicx,bm}

\makeatletter
\def\graphicscale{\twocolumn@sw{0.3}{0.4}}
\def\graphicthreescale{\twocolumn@sw{0.3}{0.4}}

\begin{document}

\title{ Scaling phenomena driven by inhomogeneous conditions 

at  first-order quantum transitions}

\author{Massimo Campostrini$^1$, Jacopo Nespolo$^1$, 
Andrea Pelissetto$^2$, and Ettore Vicari$^1$} 

\address{$^1$ Dipartimento di Fisica dell'Universit\`a di Pisa
        and INFN, Largo Pontecorvo 3, I-56127 Pisa, Italy}
\address{$^2$ Dipartimento di Fisica dell'Universit\`a di Roma ``La Sapienza"
        and INFN, Sezione di Roma I, I-00185 Roma, Italy}

\date{\today}

\begin{abstract}

We investigate the effects of smooth inhomogeneities at first-order
quantum transitions (FOQT), such as those arising from the presence of
a space-dependent external field, which smooths out the typical
discontinuities of the low-energy properties.  We argue that scaling
phenomena develop at the transition region where the external field
takes the value corresponding to the FOQT of the homogenous system.
We present numerical evidence of such scaling phenomena at the FOQTs
of quantum Ising chains, driven by a parallel magnetic field when the
system is in the ferromagnetic phase, and at the FOQT of the $q$-state
Potts chain for $q>4$, driven by an even temperature-like parameter
giving rise to a discontinuity of the ground-state energy density.

\end{abstract}


\maketitle


\section{Introduction}
\label{intro}

The theories of {\em classical} and {\em quantum} phase
transitions~\cite{Landau-book,Wilson-works,Sachdev-book} generally
apply to homogenous systems.  However, homogeneity is often an ideal
limit of experimental conditions.  Inhomogeneous conditions generally
smooth out the singularities at phase transitions.  This is also
expected at first-order transitions which are characterized by
discontinuities in the thermodynamic quantities at classical
finite-temperature transitions, or in the properties of the ground
state at first-order quantum transitions (FOQTs).

In the presence of smooth inhomogeneities, we may simultaneously
observe different phases at different space regions, separated by
crossover regions where the system passes from one phase to the other
one, developing critical correlations.  For example, this scenario is
observed in typical cold-atom experiments~\cite{BDZ-08}, where the
atoms are constrained in a limited space region by an inhomogeneous
(usually harmonic) trap, which effectively makes the chemical
potential space dependent.

The effects of the inhomogeneous conditions have been much
investigated at continuous
transitions~\cite{BDZ-08,MSGH-79,DSMS-96,WATB-04,RM-04,
  FWMGB-06,NCK-06,BKS-07,PKT-07,DZZH-07,DRBOKS-07,HK-08,
  GZHC-09,BB-09,Taylor-09,CV-09,BSB-09,ZKKT-09,RBTS-09,HR-10,
  Trotzky-etal-10,CV-10,ZH-10,HZ-10,PPS-10,PPS-10b,NNCS-10,
  ZKKT-10,CV-10b,QSS-10,FCMCW-11,ZHTGC-11,CV-11,MDKKST-11,
  HM-11,CTV-12,Pollet-12,CT-12,KLS-12,CR-12,CNPV-13,CN-14,BDV-14}.
For sufficiently smooth inhomogeneities, classical or quantum systems,
at classical (finite-temperature) or quantum (zero-temperature)
transitions, develop scaling phenomena with respect to the length
scale $\ell$ induced by the inhomogeneity. These scaling behaviors are
controlled by the universality class of the transition of the
homogenous system. They have some analogies with the standard
finite-size scaling (FSS) theory for homogenous
systems~\cite{FBJ-73,Cardy-88}, with two main differences: the
inhomogeneity due to the space-dependence of the external field
characterized by the length scale $\ell$, and a nontrivial power-law
dependence of the correlation length $\xi$ when increasing $\ell$ at
the critical point, i.e. $\xi\sim \ell^\theta$, where $\theta$ is a
universal exponent depending on some general features of the external
space-dependent field.~\cite{CV-09,CV-10}

Scaling phenomena also emerge at first-order classical transitions in
the presence of a temperature gradient~\cite{BDV-14} or a
space-dependent external field. They are observed in the transition
region where the space-dependent temperature assumes values close to
the critical temperature of the homogenous system.  The
discontinuities of the homogenous system in the thermodynamic limit
turn out to be reconstructed through scaling behaviors characterized
by nontrivial power laws, whose main features turn out to be quite
similar to those at continuous transitions.

In this paper we study the effects of inhomogeneous conditions at
FOQTs.  FOQTs are also of great interest, as they occur in a large
number of quantum many-body systems, such as quantum Hall
samples~\cite{PPBWJ-99}, itinerant ferromagnets~\cite{VBKN-99}, heavy
fermion metals~\cite{UPH-04,Pfleiderer-05,KRLF-09}, etc. They are also
expected in multicomponent cold-atom systems in optical lattices, with
spin-orbit coupling and synthetic gauge fiels, which lead to various
phases with some quantum transitions of first order, see, e.g.,
Refs.~\onlinecite{Jeckelmann-02,BRS-09,RDSG-12,PMVPFR-14,PCMS-14}.

We investigate the scaling phenomena arising when one of the model
parameters smoothly depends on the space, smoothing out the
discontinuities of the ground state.  We put forward a scaling theory
which describes the low-energy properties in the crossover space
region where the system changes phase.  We apply this scaling theory
to relatively simple quantum many-body systems, such as quantum Ising
and Potts chains driven across their FOQTs by space-dependent magnetic
fields, and check its predictions against numerical results.

The paper is organized as follows.  In Sec.~\ref{models} we present
the quantum Ising and Potts chains with external space-dependent
magnetic fields; we also show some numerical results for their
behavior around the spatial point corresponding to the parameter
values of the FOQT.  In Sec.~\ref{scalFOQT} we put forward scaling
ansatzes to describe the scaling phenomena in the crossover region
around the transition point.  In Sec.~\ref{numres} we check these
scaling theory by analyzing the numerical results of the Ising and
Potts chains. Finally, in Sec.~\ref{conclusions} we draw our
conclusions.

\section{Quantum Ising and Potts chains}
\label{models}

In order to make our scaling arguments more concrete, we first present
the quantum models that we use as theoretical laboratories for
scaling phenomena at FOQTs in the presence of a spatial
inhomogeneity. We consider the FOQTs of the Ising chains in the
ordered phase driven by a {\em parallel} magnetic field coupled to the
order-parameter spin operators, and the FOQTs of quantum $q$-state
($q>4$) Potts chains driven by a {\em transverse} magnetic field.

We also present numerical results obtained using standard
implementations of the density matrix renormalization-group (DMRG)
method~\cite{Sch-05}.  Some details of the DMRG implementations can be
found in Refs.~\onlinecite{CNPV-14,CNPV-15} where we presented
numerical analyses of the same models in homogenous conditions. The
inhomogeneous conditions that we consider here do not lead to further
particular problems from the numerical point of view.

\subsection{The quantum Ising chain}
\label{ischain}

\begin{figure}[tbp]
\includegraphics*[scale=\graphicscale]{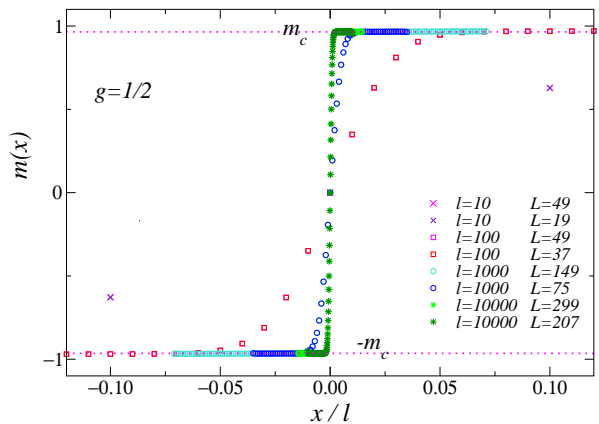}
\caption{(Color online) The local magnetization $m(x) \equiv \langle
  \sigma^{(1)}_x\rangle$ for the Ising chain (\ref{Iscjx}) with
  $g=1/2$ and $h_x=x/\ell$, versus $x/\ell$. The dotted lines indicate
  the transition values $m_\pm = \pm m_c$, cf. Eq.~(\ref{m0v}).  The
  data for the same $\ell$ and different $L$ turn out to be
  indistinguishable, showing that they effectively provide the
  $L\to\infty$ limit.  }
\label{mahxg1o2}
\end{figure}

We consider a quantum Ising chain of size $2L+1$ with a
space-dependent parallel magnetic $h_x$ along the order-parameter
spin operator, i.e.
\begin{eqnarray}
H_I &=& - J \sum_{x=-L}^{L-1} \sigma^{(1)}_x \, \sigma^{(1)}_{x+1} 
\label{Iscjx} \\
&&- g \sum_{x=-L}^L \sigma^{(3)}_x - \sum_{x=-L}^L h_x \,\sigma^{(1)}_x ,
\nonumber
\end{eqnarray}
where $\sigma^{(a)}_x$ are the Pauli matrices, $g\ge 0$ is a
transverse magnetic field, and $h_x$ is a space-dependent 
magnetic field 
\begin{equation}
h_x \equiv h(x/\ell),
\label{hx}
\end{equation}
where $\ell$ is a length scale.  The most interesting case is a linear
space dependence
\begin{equation}
h(x) = x.
\label{hxdef}
\end{equation}
Indeed, it may be also considered as a local effective linear
approximation of a more general dependence, i.e.
\begin{equation}
h(x) \approx a_1 (x -x_0) +a_2 (x-x_0)^2 + ....  
\label{hxgen}
\end{equation}
around the point $x_0$ where it vanishes, corresponding to the FOQT
value.  It is also convenient to extend the analysis to a more general
power law of the space dependence
\begin{equation}
h(x)= {\rm sgn}(x) \; |x|^{p},
\label{hip}
\end{equation}
to crosscheck the scaling theory we shall put forward to describe
these phenomena.  We study the system in the $L\to\infty$ limit and
investigate the scaling behavior with respect to the remaining length
scale $\ell$.  In the following we set $J=1$.

Note that the $p\to\infty$ limit of Eq.~(\ref{hip}) corresponds to a
homogenous system with $L=\ell$ and fixed opposite (kink-like)
boundary conditions (FOBC), which may be described by the standard
Ising-chain Hamiltonian with a boundary term~\cite{CNPV-14}
\begin{eqnarray}
H_{I,{\rm FOBC}} &=& - \sum_{x=-L}^{L-1} \sigma^{(1)}_x \sigma^{(1)}_{x+1} 
\label{Iscfobc}\\
&&- g \sum_{x=-L}^L \sigma^{(3)}_x  - (\sigma_{-L}^{(1)} - \sigma_L^{(1)}).
\nonumber
\end{eqnarray}
The last term between parenthesis is added to achieve FOBC, indeed it
arises when adding further fictitious sites at $-L-1$ and $L+1$ which
are eigenstates of $\sigma^{(1)}$ with opposite $\pm 1$ eigenvalues
respectively.

The homogenous Ising chain, i.e. the model (\ref{Iscjx}) with constant
magnetic field $h_x=h$, has a continuous transition at $g=1, \, h=0$,
belonging to the two-dimensional Ising universality class.  This
quantum critical point separates a paramagnetic ($g>1$) and a
ferromagnetic ($g<1$) phase.  In the ferromagnetic phase $g<1$, the
parallel magnetic field $h$ drives a FOQT at $h=0$, with a
discontinuity of the magnetization, i.e. the ground-state expectation
value of $\sigma^{(1)}_x$.  Indeed, neglecting boundary effects, we
have~\cite{Pfeuty-70}
\begin{eqnarray}
&&m_\pm = {\rm lim}_{h\to 0^\pm} {\rm lim}_{L\to\infty} 
\langle \sigma^{(1)}_x \rangle = \pm m_c, \label{mpm}\\
&& m_c=(1-g^2)^{1/8}.
\label{m0v}
\end{eqnarray}
Therefore, in the presence of an inhomogeneous field which vanishes at
$x=0$ changing sign, such as that in Eq.~(\ref{hxdef}), the point
$x=0$ effectively corresponds to a spatial transition point where the
system experiences a transition between two magnetized phases with
opposite sign, i.e. with 
$\langle \uparrow | \sigma^{(1)}_x |\uparrow\rangle = m_+$ and
$\langle \downarrow | \sigma^{(1)}_x |\downarrow\rangle = m_-$.

\begin{figure}[tbp]
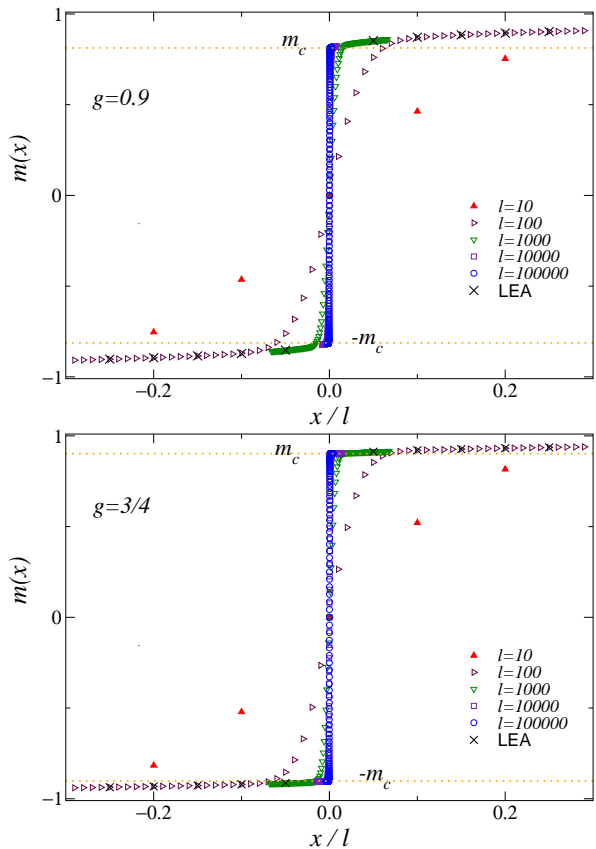

\includegraphics*[scale=\graphicscale]{sxplp1g0p9.eps}
\includegraphics*[scale=\graphicscale]{sxplp1g.eps}
\caption{(Color online) The local magnetization $m(x)$ of the model
  (\ref{Iscjx}) with $h_x=x/\ell$, for $g=3/4$ (bottom) and $g=9/10$
  (top).  The values of $m_c$ are given by Eq.~(\ref{m0v}).  We also
  show data computed using the LEA, obtained by DMRG computations of
  the homogenous systems.}
\label{mahxotherg}
\end{figure}

\begin{figure}[tbp]
\includegraphics*[scale=\graphicscale]{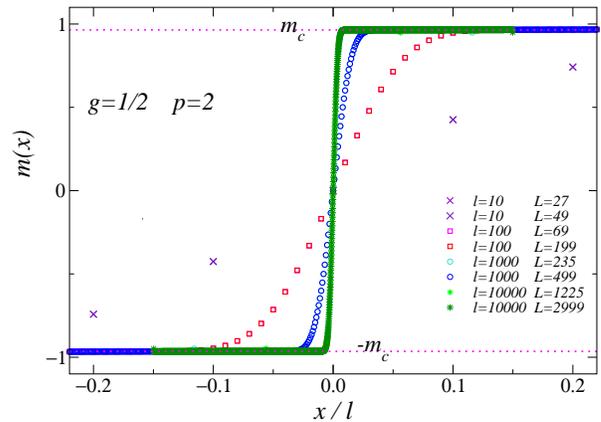}
\caption{(Color online) The local magnetization $m(x)$ for the
  Hamiltonian (\ref{Iscjx}) with quadratic space dependence ($p=2$) of
  $h_x$ at $g=1/2$.  The data for the same $\ell$ and different $L$
  practically coincide, showing that they are already a good
  approximation of the $L\to\infty$ limit.}
\label{mahxp2}
\end{figure}

Some numerical DMRG results for the space dependence of the local
magnetization
\begin{equation}
m(x) = \langle \sigma^{(1)}_x\rangle
\label{mxdef}
\end{equation}
are shown in Figs.~\ref{mahxg1o2}, \ref{mahxotherg} and \ref{mahxp2}
for linearly and quadratically varying fields $h(x)$.  They are
obtained for sufficiently large size $L$, to effectively provide their
$L\to\infty$ limit at fixed $\ell$ around the region where $h(x)$
vanishes.  This is easily checked by comparing data with the same
$\ell$ and increasing $L$; some examples are reported in
Figs.~\ref{mahxg1o2} and \ref{mahxp2}.  Note that, with increasing
$\ell$, we need smaller and smaller ratios $L/\ell$ to achieve the
large $L$ limit for the energy differences of the lowest levels, and
the observables around $x=0$. This will be explained by the scaling
theory of Sec.~\ref{scalFOQT}, which shows that the relevant scaling
length in the crossover region around $x=0$ is $\xi\sim \ell^{\theta}$
with $\theta=1/4$ for linear $h(x)$, and $\theta=2/5$ for the
quadratic dependence. Therefore, the relevant ratio for the crossover
region is $L/\ell^\theta$, instead of $L/\ell$.

The data around $x=0$ show a crossover between the two magnetization
values $m_\pm$, cf. Eq.~(\ref{mpm}), which becomes sharper and sharper
with increasing $\ell$. The comparison between linear and quadratic
dependences of $h_x$, see in particular Figs.~\ref{mahxg1o2} and
\ref{mahxp2} for the same value $g=1/2$, show similar behaviors, only
the crossover region appears enlarged.

In noncritical regimes, away from phase transitions when correlations
do not develop long length scales, inhomogeneity effects can be
effectively taken into account by local-equilibrium approximations
(LEA), assuming a local equilibrium analogous to that of the
homogenous system at the same fixed parameters.  An example is the
local-density approximation widely used to study particle systems with
an effective space-dependent chemical potential, see, e.g.,
Refs~\onlinecite{BDZ-08,GPS-08,Esslinger-10,CV-10b,GBL-13,ACV-14}.
However, when correlations develop large length scales, such as at
classical and quantum transitions, LEA may not provide a satisfactory
description, and significant corrections are
found~\cite{CV-10b,CTV-12,ACV-14}. This failing of the LEA is also
observed at first-order classical transition in the presence of a
temperature gradient~\cite{BDV-14}.

We compare the results for the inhomogeneous Ising model with the LEA
$m_{\rm lea}(x)$, which estimates $m(x)$ using the corresponding
values $m_h(h)$ of the homogenous system in the infinite volume limit
at the given value of $h_x$, i.e.
\begin{eqnarray}
m(x) \approx m_{\rm lea}(x/\ell) =m_h[h(x/\ell)]
\label{mlta}
\end{eqnarray}
Note that since the external field $h_x$ is a function of the ratio
$x/\ell$, the LEA scales as $x/\ell$.  LEA is expected to provide a
good approximation when $h_x$ varies smoothly, thus for large $\ell$.
However, since the magnetization value of the homogenous system lies
within $1\ge |m|\ge m_c$, LEA can not describe the crossover region
where $|m|<m_c$.

Some LEA results are shown in Fig.~\ref{mahxotherg}.  The data at
fixed $x/\ell$ appear to approach their LEA with increasing $\ell$.
This convergence is fast far from $x=0$, but it becomes significantly
slower when approaching $x=0$, i.e. it is non uniform when $|x|\to
0$. As we shall see, this reflects a hidden nontrivial scaling
behavior which characterizes the crossover region around $x=0$ in the
smooth $\ell\to\infty$ limit, and requires nontrivial power-law
rescalings of the distances from $x=0$.  This is a novel regime,
somehow probing the mixed quantum phase where $-m_c < m(x) < m_c$.

\subsection{The quantum Potts chain}
\label{pottschain}

\begin{figure}
\centering \includegraphics{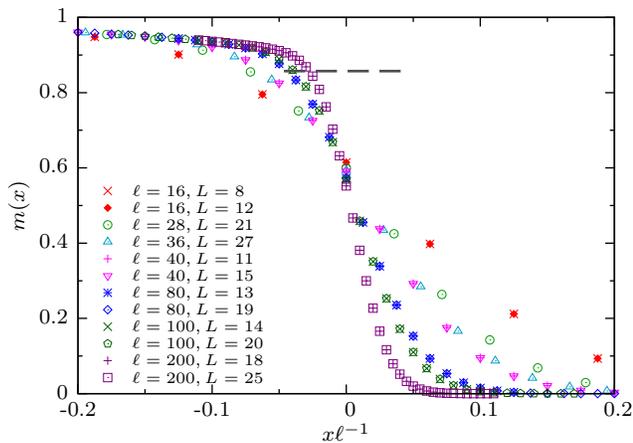}
 \caption{The local magnetization $m(x)$ of the quantum $q=10$ Potts
   chain in the presence of a linear transverse field.  The dashed
   line indicates the value $m_c=0.8572$ defined in Eq.~(\ref{m0def}).
   The data for the same $\ell$ and different $L$ practically
   coincide, showing that they are already asymptotic and do not
   depend on $L$.}
\label{mxpotts}
\end{figure}

\begin{figure}
\centering \includegraphics{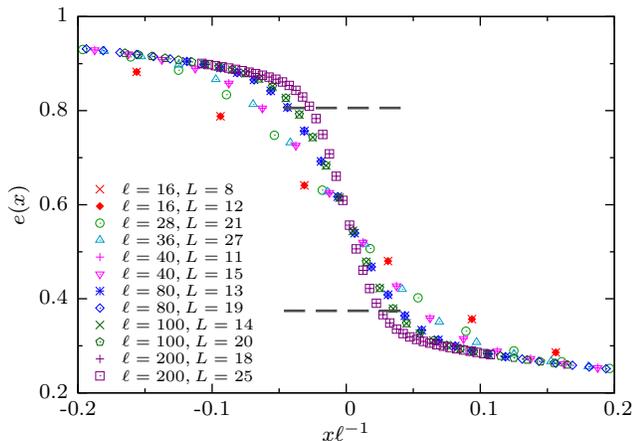}
 \caption{The energy density $e(x)$, cf. Eq.~(\ref{energydens}), of
   the quantum $q=10$ Potts chain in the presence of a linear
   transverse field.  The dashed lines indicate $e_-= 0.8060$ and
   $e_+=0.3745$ defined in Eq.~(\ref{epmdef}).  The data for the same
   $\ell$ and different $L$ practically coincide, showing that they
   are already asymptotic and do not depend on $L$.}
\label{expotts}
\end{figure}

Examples of FOQTs driven by even temperature-like parameters are
provided by the quantum $q$-state Potts chains for $q>4$, which are
the quantum counterpart of the classical two-dimensional Potts
models~\cite{Potts-52,Wu-82,Baxter-73}
\begin{equation}
H_c = - J \sum_{\langle ij\rangle } \delta(s_i,s_j), 
\label{cPotts}
\end{equation}
where the sum is over the nearest-neighbor sites of a square lattice,
$s_i$ are spin variables taking $q$ integer values, i.e.
$s_i=1,...,q$, and $\delta(m,n)=1$ if $m=n$ and zero otherwise.  The
quantum Hamiltonian can be derived from the {\em time} continuum limit
of the transfer matrix, with $q$ states per site, which can be labeled
by an integer number $|n=1\rangle,...,|n=q\rangle$.  For a chain of
size $2L+1$ it reads~\cite{SP-81,IS-83,CNPV-15}
\begin{eqnarray}
H_P = - \sum_{x=-L}^{L-1} \sum_{k=1}^{q-1} 
\Omega_x^k \Omega_{x+1}^{q-k} -  g \sum_{x=-L}^L \sum_{k=1}^{q-1} M_x^k
- \sum_{k=1}^{q-1} \Omega_{-L}^k \quad
\label{qPotts}
\end{eqnarray}
where $\Omega_x$ and $M_x$ are $q\times q$ matrices:
\begin{align}
 &\Omega = \delta_{m,n} \, \omega^{n-1},\qquad \omega = e^{i 2 \pi/q},
 \label{omega}\\
 &M = \delta_{m,{\rm Mod}(n-1,q)}=
\begin{pmatrix}
       0 & 1      &       &  \\
         & \ddots &\ddots  &   \\
         &       &        & 1 \\
       1 &        &        & 0 \\
      \end{pmatrix}.
\end{align}
These matrices commute on different sites and satisfy the algebra:
$\Omega_x^k \Omega_x^l = \Omega_x^{k+l}$, $M_x^k M_x^l = M_x^{k+l}$, $
\Omega_x^q = M_x^q = \mathbb{I}$, $M_x^k \Omega_x^l = \omega^{kl}
\Omega_x^l M_x^k$.  

Last term in the r.h.s. of Eq.~(\ref{qPotts}) is a boundary term which
softly breaks the $q$-state symmetry favoring the state $n=1$.  It
ensures the self-dual property~\cite{IS-83}
\begin{equation}
H_P(g) = g H_P(1/g)
\label{duality}
\end{equation}
even for finite chains~\cite{CNPV-15}.  The Hamiltonian $H_P$
corresponds to a chain with mixed self-dual boundary conditions
(SDBC), with a fixed state $n=1$ at a further site $x=-L-1$, and an
unmagnetized disordered state $\propto \sum_{n=1}^q |n\rangle$ at
$x=L+1$.

In the case of $q=2$ states the Hamiltonian $H_P$ describes a quantum
Ising chain with mixed fixed-free boundary conditions.

Like quantum Ising chains, the low-energy properties of the quantum
Potts chains show two phases: a disordered phase for sufficiently
large values of $g$ and an ordered phase for small $g$ where the
system magnetizes along one of the $q$ {\em directions}.  The
transition point is easily inferred from the duality relation
(\ref{duality}), obtaining $g=g_c=1$. For $q>4$ the two phases are
separated by a FOQT where the energy density and magnetization are
discontinuous~\cite{SP-81,IS-83,Baxter-73,Wu-82}.

The FOQTs of the Potts chains are characterized by a discontinuity of
the energy density of the ground state.  We define the energy density
as
\begin{equation}
e(x_b) = \langle {\cal E}_x \rangle,\quad 
{\cal E}_x = \delta(n_x,n_{x+1}) = 
{1\over q} \sum_{k=1}^q \Omega_x^k \Omega_{x+1}^{q-k},
\label{energydens}
\end{equation}
where $x_b=x+1/2$ is the position of the bond center.  The
infinite-volume energy density changes discontinuously across the
FOQT, i.e. the two limits
\begin{equation}
e_\pm = {\rm lim}_{g\to 1^\pm} \; {\rm lim}_{L\to\infty} \; e(x)
\label{epmdef}
\end{equation}
differ at the FOQTs of the Potts chains with $q>4$.  Their difference
$\Delta e\equiv e_+-e_-$ is the analog of the latent heat of
first-order finite-temperature transitions.  For
example,~\cite{CNPV-15} for the $q=10$ Potts chain $e_-= 0.8060(1)$
and $e_+=0.3745(5)$.

Also the magnetization is discontinuous at the transition, passing
from zero in the disorder ($g>1$) phase to nonzero in the ordered
($g<1$) phase.  We define the local magnetization of the ground state
as
\begin{eqnarray}
&&m(L,g,x) = \langle {\cal M}_x \rangle,\label{mlxdef}\\
&&{\cal M}_x= {q \delta(n_x,1) - 1\over q-1},
\quad \delta(n_x,1) = 
{1\over q} \sum_{k=1}^{q} \Omega_x^k.
\label{mxdefpo}
\end{eqnarray}
The limit
\begin{equation}
m_c = {\rm lim}_{g\to 1^-} \; {\rm lim}_{b\to 0}  
\; {\rm lim}_{L\to\infty} \; m(x)
\label{m0def}
\end{equation}
is non zero for $q>4$, where $b$ is a {\em magnetic} field coupled to
the global projector to the $n=1$ state, e.g. described by the
Hamiltonian term
\begin{equation}
H_{Pb} = - b \, \sum_{x=-L}^L \sum_{k=1}^{q} \Omega_x^k.
\label{hgpotts}
\end{equation}
For example, numerical results for $q=10$ give~\cite{CNPV-15}
$m_c=0.8572(1)$.

Again, we extend the homogenous model (\ref{qPotts}) to allow for a
space-dependent {\em transverse magnetic} field. This is achieved by
adding
\begin{eqnarray}
H_{Ph} = -  \sum_{x=-L}^L h_x \sum_{k=1}^{q-1} M_x^k
\label{hhpotts}
\end{eqnarray}
to the Hamiltonian (\ref{qPotts}), where $h_x$ may have a linear space
dependence such as Eq.~(\ref{hxdef}), or a more general power law such
as Eq.~(\ref{hip}).  We fix the parameter $g$ of the Hamiltonian $H_P$
to its critical value $g=g_c=1$, so that at the center $x=0$ of the
chain the parameters take the values of the FOQT. Moreover, we
consider $L\le \ell$ so that the local transverse field satisfies
$g+h_x>0$.  Again when we consider external fields (\ref{hip}) in the
limit $p\to\infty$, we recover the homogenous system with SDBC.

In Fig.~\ref{mxpotts} we show DMRG results for the local magnetization
$m(x)$ of the $q$-state Potts chain with $q=10$ in the presence of a
linearly space-dependent field $h_x$, cf. Eq.~(\ref{hxdef}).  They
show that the local magnetization rapidly drops in the space region
corresponding to the disordered phase, i.e. $x>0$.  

Data for the energy density, and its space dependence, are shown in
Fig.~\ref{expotts}. They clearly show a crossover region where the
data pass from $e(x)\gtrsim e_-$ to $e(x)\lesssim e_+$, which are the
values of the energy density corresponding to the ordered and
disordered phase respectively.

The data of the energy differences of the lowest states, and the
observables around $x=0$, i.e. for sufficiently small ratios $x/\ell$,
rapidly converge when increasing the ratio $L/\ell$ keeping $\ell$
fixed.  Like the Ising case, this is checked by comparing data with
increasing $L$, as shown in Figs.~\ref{mxpotts} and \ref{expotts}.
For example, in the case of $\ell=200$, $E_1-E_0=0.90956$ for $L=18$
and $E_1-E_0=0.90952$ for $L=25$.  Analogous precision is achieved for
the other observables around $x=0$, and sufficiently far from the
boundaries.  Again, with increasing $\ell$ smaller and smaller ratios
$L/\ell$ are sufficient to effectively obtain $L$-independent results.
This is essentially related to the fact the relevant scaling length in
the crossover region around $x=0$ is $\xi\sim \ell^\theta$ with
$\theta=1/3$ for linear $h(x)$, as argued in Sec.~\ref{scalFOQT}.

Note that, since the Potts chain with $q=10$ is much more complex than
the Ising chain, DMRG computations allow us to get reliable results
for smaller chain sizes, and therefore smaller length scales of the 
external magnetic field. The is essentially related to the fact that
many more states per site must be kept in the computations.

\section{Scaling behavior at the crossover space region}
\label{scalFOQT}

In this section we present a scaling theory for the behaviors observed
at the FOQTs of the Ising and Potts chains in the presence of
inhomogeneous external fields.

For this purpose we first consider the $p\to\infty$ limit of the
external field (\ref{hip}), which corresponds to homogenous systems of
finite size with appropriate boundary conditions.  In the case of the
FOQTs of the Ising chains the resulting boundary conditions are FOBC,
see Sec.~\ref{ischain} and in particular Eq.~(\ref{Iscfobc}).  In the
case of the FOQTs of the Potts chain, see Sec.~\ref{pottschain}, the
$p\to\infty$ limit corresponds to the homogenous system with SDBC,
i.e. Eq.~(\ref{qPotts}) with $h_x=0$.

Therefore, in the $p\to\infty$ limit the scaling behavior must match
the finite-size behavior of homogeous systems at FOQTs.  Although
FOQTs do not develop a diverging correlation length in the
infinite-volume limit, they show FSS behaviors around the transition
point, both in the case of classical and quantum first-order
transitions~\cite{NN-75,FB-82,PF-83,IS-83,FP-85,CLB-86,Privman-90,
  LK-91,BK-92,BNB-93,VRSB-93,IC-99,CPPV-04,CNPV-14,CNPV-15}.  The FSS at
FOQTs turns out to be particularly sensitive to the boundary
conditions.  Indeed, the size dependence of the scaling variables may
significantly change when varying the boundary
conditions.~\cite{CNPV-14,CNPV-15} For example, in the case of the
FOQTs of Ising chains, driven by a magnetic field in their ordered
quantum phase, we have an exponential size dependence for open and
periodic boundary conditions, while it is power law for antiperiodic
or kink-like FOBC boundary conditions.~\cite{CNPV-14} Actually, this
particular sensitiveness to the boundary conditions is a peculiar
feature of FOQTs, which qualitatively distinguish their FSS behaviors
from those at continuous quantum transitions, see
e.g. Refs.~\onlinecite{SGCS-97,CPV-14}.

The relevant {\em scaling} variable of FSS at FOQTs is given by the
ratio $\kappa=E_L/\Delta_L$ between the energy contribution $E_L$ of
the perturbation driving the transition and the energy difference
({\em gap}) of the lowest states $\Delta_L\equiv E_1-E_0$ at the
transition point.  The particular sensitiveness to the boundary
conditions essentially arises from the gap $\Delta_L$ entering the
scaling variable $\kappa$, whose finite-size behavior depends
crucially on the boundary conditions considered.  At the FOQTs ($h=0$)
of the Ising chain with FOBC the gap behaves as~\cite{CNPV-14}
\begin{equation}
\Delta_L\sim L^{-z}, \qquad z=2,
\label{delis}
\end{equation}
which may be associated with a dynamic exponent $z=2$.  In the case of
FOQTs of the Potts chain with SDBC it behaves as~\cite{CNPV-15}
\begin{equation}
\Delta_L\sim L^{-z},\qquad z=1,
\label{delpo}
\end{equation}  
thus corresponding to a dynamic exponent $z=1$.

At the FOQTs of the Ising chain driven by the parallel magnetic field
$h$, the relevant scaling variable of its FSS with FOBC
is~\cite{CNPV-14}
\begin{equation}
\kappa_I = h L/\Delta_L \sim h L^{d+z} = h L^3.
\label{kappaisi}
\end{equation}
In the language of the renormalization-group (RG) theory, this
relation allows us to associate a RG dimension with the perturbation
$h$, given by
\begin{equation}
y_h=d+z=3.
\label{yhisi}
\end{equation}

The FOQTs of the Potts chains for $q>4$ is driven by the model
parameter $g$.  Setting the perturbation $h\equiv g-1$ at the
transition point $g=1$, the relevant scaling variable for SDBC
turns out to be~\cite{CNPV-15}
\begin{equation}
\kappa_P = h L/\Delta_L \sim h L^{d+z} = h L^2.
\label{kappapo}
\end{equation}
Thus 
\begin{equation}
y_h=d+z=2
\label{yhpotts}
\end{equation}
is the RG dimension of $h=g-1$ describing the FSS at
the FOQTs of the Potts chains with SDBC.  

The above considerations imply that the space dependence is controlled
by the scaling variable
\begin{equation}
x/L \sim x h^{1/y_h}
\label{xsca}
\end{equation}
to keep $\kappa_{I,P}$ fixed.  We want to extend the FSS
ansatzes~\cite{CNPV-14} holding for the homogenous systems, thus in
the limit $p\to\infty$, to allow for a power-law space dependence of
the external fields.  The scaling variables in the presence of
inhomogenous fields characterized by the power law $p$,
cf. Eq.~(\ref{hip}), can be heuristically derived by replacing the
perturbation parameter $h$ with $h_x \sim (x/\ell)^p$ in
Eq.~(\ref{xsca}).  Therefore, assuming that the scaling behavior
remains controlled by the RG dimension $y_h$, and that the chain size
$L$ is sufficiently large not to play any role, we obtain
\begin{equation}
x \,\left({x\over \ell}\right)^{p/y_h} = 
\left( {x\over \ell^\theta} \right)^{1 + p/y_h},
\label{xscader}
\end{equation}
where the exponent $\theta$ is given by
\begin{equation}
\theta = {p\over p+y_h},
\label{theta}
\end{equation}
with $y_h$ given by Eqs.~(\ref{yhisi}) and (\ref{yhpotts})
for the FOQTs of Ising and Potts chains respectively.
The relation (\ref{xscader}) suggests that the relevant scaling in the
presence of inhomogeneous external fields is obtained by keeping the
scaling variable
\begin{equation}
X = x/\ell^\theta
\label{ysca}
\end{equation}
fixed.  This implies that the observables and correlations in the
crossover region around the transition point develop a length scale
$\xi$, behaving as
\begin{equation}
\xi\sim \ell^\theta.
\label{xidef}
\end{equation}
Note that $\theta\to 1$ for $p\to \infty$ consistently with the fact
that we must recover the FSS of homogenous systems in this limit.

On the basis of these considerations we expect that the asymptotic
large-$\ell$ behavior of the energy difference $\Delta_\ell$ of the
two lowest levels scales as
\begin{equation}
\Delta_\ell \sim \xi^{-z} \sim \ell^{-z\theta}
\label{deltasca}
\end{equation}
with $\theta$ given by Eq.~(\ref{theta}), and $z$ is the effective
dynamic exponent read from the size dependence of the gap at the
transition point.

Around the point where $h(x)$ vanishes, the local magnetization is
expected to asymptotically behave as
\begin{eqnarray}
m(x) =  m_c \, f_m(X),\qquad X=x/\ell^\theta,
\label{magsccl}
\end{eqnarray}
where $m_c$, cf. Eqs.~(\ref{mpm}) and (\ref{m0def}), is the
normalization such that $\lim_{X\to -\infty} f_m(X)=-1$ in the Ising
case, and $\lim_{X\to -\infty} f_m(X)=1$ in the Potts case.  We also
consider the two-point function of the order parameter.  In the case
of the Ising chain it is defined as
\begin{eqnarray}
G(x,y) = \langle \sigma_{x}^{(1)} \sigma_{y}^{(1)} \rangle.
\label{twopfi}
\end{eqnarray}
In the case of the Potts chain we consider the two-point correlation
function
\begin{eqnarray}
G(x,y) = \langle {\cal M}_{x} {\cal M}_{y} \rangle,
\label{twopfp}
\end{eqnarray}
and its connected part  
\begin{equation}
G_c(x,y) = \langle {\cal M}_{x} {\cal M}_{y} \rangle -
\langle {\cal M}_{x} \rangle \langle {\cal M}_{y} \rangle,
\label{gconn}
\end{equation}
with ${\cal M}_x$ defined in Eq.~(\ref{mxdefpo}).  Around the region
where $h$ vanishes, we expect the scaling behavior
\begin{eqnarray}
G(x_1,x_2) \approx m_c^2 \, f_g(X_1,X_2).
\label{mp1o}
\end{eqnarray}
An analogous scaling is expected for its connected part $G_c(x_1,x_2)$.
The scaling functions $f_m$ and $f_g$ are expected to be universal,
i.e. largely independent of the microscopic details of the model.  For
example, in the case of the FOQT of Ising chains, they are expected to
be independent of the particular value of $g$ within the quantum
ordered phase, apart from a trivial (and unique) rescaling of their
arguments.

When the FOQT gives rise to a discontinuity in the energy density,
such as the FOQT of quantum Potts chains with $q>4$ at $g=1$, we
expect that its asymptotic behavior around $x=0$ is
\begin{equation}
e(x) \approx f_e(X). 
\label{enesca}
\end{equation}
Moreover, the scaling function $f_e$ is expected to have the
value $e_\pm$, cf. Eq.~(\ref{epmdef}), as asymptotic limits, i.e.
\begin{equation}
{\rm lim}_{X\to\pm\infty} f_e(X) = e_\pm.
\label{limxfe}
\end{equation}
essentially because it describes the crossover between the two pure
phases where the energy density of the system takes the values
$e_\pm$.

The above large-$\ell$ scaling ansatzes are expected to be approached
with $O(\ell^{-\theta})$ corrections.  Note that they also imply that
the curves for different values of $\ell$ cross each other around
$x=0$, as shown in Figs.~\ref{mahxg1o2}-\ref{expotts}, and this
crossing point approaches the point $X_c$ (where $X=x/\ell$)
corresponding to $g=g_c$.  Actually, one may exploit this property to
estimate the critical parameter $g_c$ when it is not known, using a
linear spatial dependence of $g$ (for which the point $x=0$ is not
particular) and looking at the crossing point of the energy density
and magnetization data.  The results are expected to approach $X_c$,
thus $g_c$, with $O(\ell^{-1})$ corrections.

It is important to note that, in the case of more general space
dependences such as that in Eq.~(\ref{hxgen}), the linear term
determines the scaling behavior at the crossover region, obtained
keeping $X\equiv x/\ell^\theta$ fixed with $\theta=(1+y_h)^{-1}$,
cf. Eq.~(\ref{theta}) with $p=1$, while higher-order terms give rise
to $O(\ell^{-1+\theta})$ corrections.

Although the above discussion focuses on the FOQTs of the Ising and
Potts chains, the scaling ansatzes at FOQTs in the presence of
inhomogeneous fields can be straightforwardly extended to general
FOQTs, and higher dimensions.

Finally, we note that similar scaling behaviors have been conjectured,
and numerically checked, at classical first-order transitions in the
presence of a temperature gradient~\cite{BDV-14}.

\section{Scaling phenomena induced by the inhomogeneous fields}
\label{numres}

In this section we show that the numerical results for the Ising and
Potts chains in the presence of inhomogeneous magnetic fields support
the scaling behaviors put forward in Sec.~\ref{scalFOQT}.  We study
the scaling behavior with respect to the length scale $\ell$ only; as
already discussed in Sec~\ref{models}, the data that we present are
obtained for sufficiently large $L$, so that their behaviors in the
crossover region do not effectively depend on $L$ anymore.

\subsection{Results for the Ising chain}
\label{numisch}

\begin{figure}[tbp]
\includegraphics*[scale=\graphicscale]{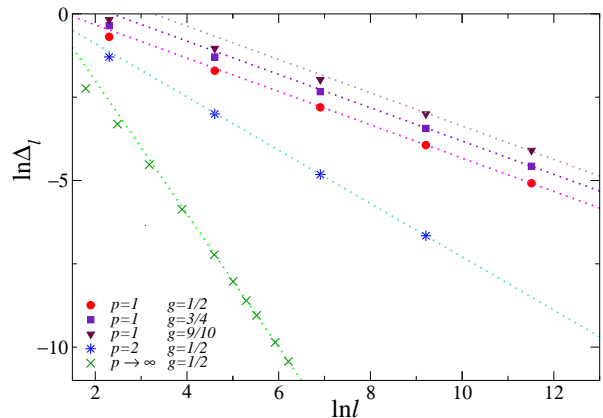}
\caption{(Color online) The gap $\Delta_\ell$ as a function of $\ell$
  for the Hamiltonian (\ref{Iscjx}), for magnetic fields $h_x$ with
  power laws $p=1$, $p=2$ and $p\to\infty$.  The dotted lines show the
  expected behavior $\Delta\sim \ell^{-2\theta}$ with $\theta=1/4$ for
  $p=1$, $\theta=2/5$ for $p=2$, and $\theta=1$ for $p=\infty$.  }
\label{dehx}
\end{figure}

\begin{figure}[tbp]
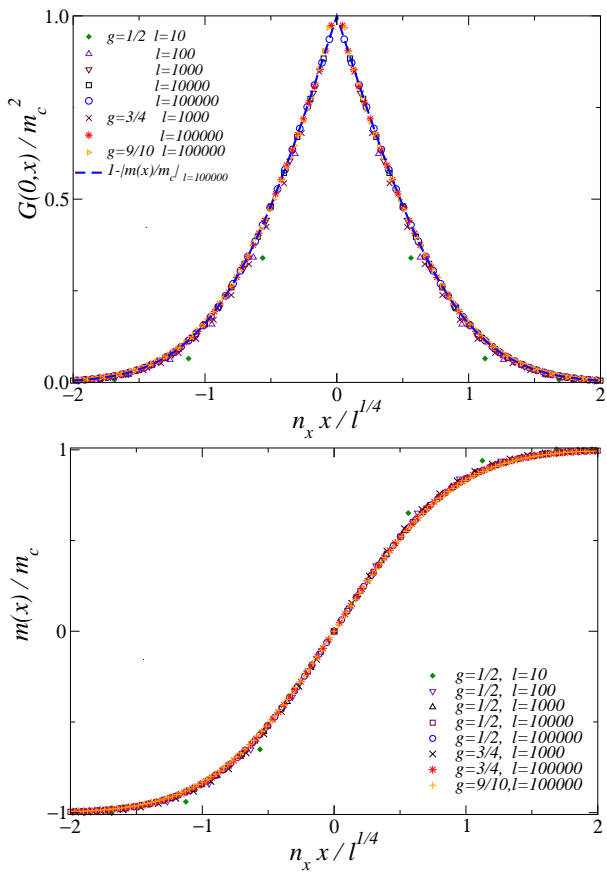

\includegraphics*[scale=\graphicscale]{gxrp1g.eps}
\includegraphics*[scale=\graphicscale]{sxrp1g.eps}
\caption{(Color online) Scaling of the local magnetization $m(x)$ and
  the two-point function $G(0,x)$, for the Hamiltonian (\ref{Iscjx})
  at $g=1/2,\,3/4,\,9/10$, with a linear magnetic field,
  cf. Eq.~(\ref{hxdef}). We plot the ratios $m(x)/m_0$ (bottom) and
  $G(0,x)/m_c^2$ (top) versus $n_x x/\ell^{1/4}$, where $n_x$ is a
  normalization, in the figure we use $n_{x}\approx 1,\,3/4,\,4/7$ and
  for $g=1/2,\,3/4,\,9/10$ respectively.  In the top figure, the
  comparison with the data of of $1-|m(x)/m_c|$ supports the
  prediction (\ref{goxkink}) of the one-kink scenario.  }
\label{mgscaxp1}
\end{figure}

Fig.~\ref{dehx} shows the dependence on the length scale $\ell$ of the
energy difference $\Delta_\ell$ of the lowest states for various
values of $p$, i.e. $p=1,\,2$ and $p\to\infty$ corresponding to
homogenous systems with FOBC. They confirm the predicted behavior
$\Delta_\ell\sim \ell^{-2\theta}$, cf. Eq.~(\ref{deltasca}) with
$z=2$.

In Fig.~\ref{mgscaxp1} we show results for the local magnetization
$m(x)$ and the two-point function $G(0,x)$ in the case of a linear
dependence ($p=1$) of $h_x$, for which $\theta=1/4$, at three values
of $g$ to check universality, i.e. $g=1/2,\;3/4,\;9/10$.  They nicely
confirm the asymptotic scaling behavior predicted by the
Eqs.~(\ref{magsccl}) and (\ref{mp1o}), and the universality of the
scaling functions $f_m$ and $f_g$ with respect to $g$, apart from a
trivial rescaling of its argument.

Like the homogenous system with kink-like FOBC, we expect the lowest
energy states are associated with domain walls (kinks), i.e. nearest
neighbors pairs of antiparallel spins, which can be considered as
one-particle states. In homogenous systems~\cite{CJ-87,CNPV-14} they
have $O(L^{-1})$ momenta, giving rise to a gap of order $L^{-2}$ for
FOBC.  We expect an analogous scenario for the ground state in the
presence of the linear magnetic field $h_x=x/\ell$, that is the ground
state is a superposition of one-kink states which switch the chain
sites from $|\downarrow\rangle$ to $|\uparrow\rangle$.  In particular
we argue that this picture describes the crossover region described by
the scaling ansatzes (\ref{magsccl}) and (\ref{mp1o}), which
interpolates between the states with magnetization $m_\pm$,
cf. Eq.~(\ref{mpm}). In this one-kink scenario the local magnetization
$m(x)$ and the two-point function $G(0,x)$ must be asymptotically
related.  If we define $p(x_1,x_2)$ the probability to find the kink
in the interval $(x_1,x_2)$, then the scaling function $f_m$ is
\begin{equation}
f_m(X) = {m(x)\over m_c} = 2 p(-\infty,X) - 1,
\label{mrx}
\end{equation}
where $X\equiv x/\ell^\theta$, and $m_c$ is the infinite-volume
  magnetization, which provides the normalization of the scaling
  relation (\ref{magsccl}).  Also the value of the two-point function
  $G(0,x)$ is related to the probability to find the kink in the
  region $(0,x)$, i.e.
\begin{equation}
f_g(0,X) = {G(0,x)\over m_c^2} = 1 - p(0,X).\label{gpro}
\end{equation}
Since $p(\infty,0)=1/2$ by symmetry,
\begin{equation}
p(0,X) = p(-\infty,X) - p(-\infty,0)=p(-\infty,X)-1/2
\label{q0x}
\end{equation}
for $X>0$, and the analog for $X<0$. Thus we obtain the relation
\begin{equation}
f_g(0,X) = 1 - |f_m(X)|.
\label{goxkink}
\end{equation}
This relation is confirmed by the data, see the top
Fig.~\ref{mgscaxp1}.

\begin{figure}[tbp]
\includegraphics*[scale=\graphicscale]{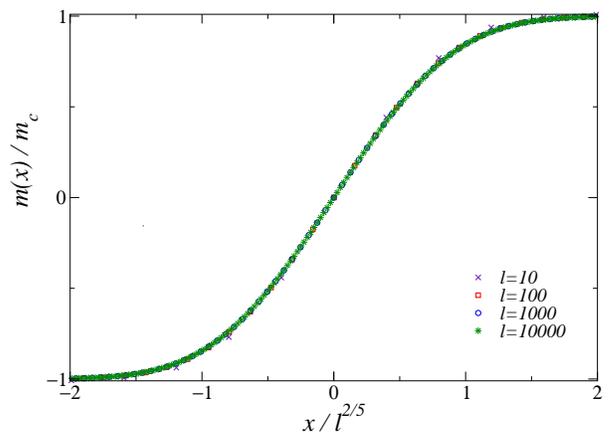}
\caption{(Color online) $m(x)/m_c$ 
  versus $x/\ell^{2/5}$, for the Hamiltonian (\ref{Iscjx}) at
  $g=1/2$ with a quadratic magnetic field, cf. Eq.~(\ref{hip}) with
  $p=2$.  Like the linear case $p=1$, the asymptotic one-kink relation 
  (\ref{goxkink}) is satisfied by the data.
  }
\label{mgscaxp2}
\end{figure}

Analogous results are obtained in the case of quadratic dependence,
i.e.  $p=2$ in Eq.~(\ref{hip}), with $\theta=2/5$, see
Fig.~\ref{mgscaxp2}.  As already mentioned, the scaling behaviors in
the $p\to\infty$ limit must reproduce the FSS of the Ising chain with
FOBC.~\cite{CNPV-14} In particular, for any $g<1$ and $h=0$, the FSS
functions of the local magnetization and two-point function are given
by~\cite{CPV-15}
\begin{eqnarray}
&& f_m(X) = 
 X + {1\over \pi} \sin(\pi X), \qquad X = x/\ell,
\label{mxpinf}\\
&&f_g(X_1,X_2) = 1 - |f_m(X_2)-f_m(X_1)| 
\label{goxkinkpi}
\end{eqnarray}
in the large-$\ell$ limit keeping $X$ fixed, with $-1\le X  \le
1$.

\subsection{Results for the $q=10$ Potts chain}
\label{numpoch}

We now present an analogous analysis of the DMRG data of the $q=10$
Potts chain with a linearly varying field $h_x$,
cf. Eq.~(\ref{hhpotts}).  In this case we have that $\theta=1/3$
according to Eq.~(\ref{theta}).

\begin{figure}
\centering \includegraphics{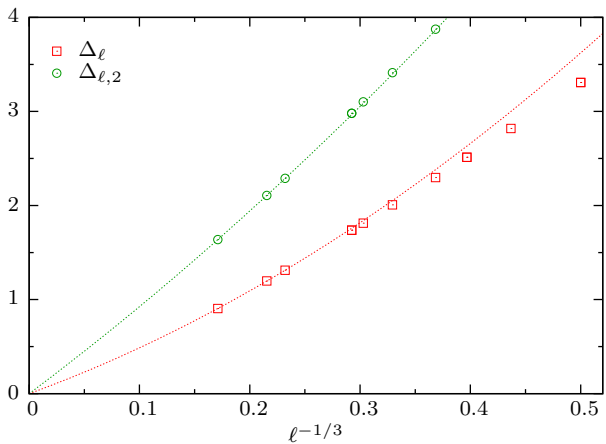}
 \caption{The  $\ell$-dependence of the energy differences of the lowest
   states, i.e. $\Delta_\ell = E_1-E_0$ and $\Delta_{\ell,2}=E_2-E_0$.
   They are consistent with an asymptotic $\ell^{-1/3}$ suppression,
   as predicted by Eq.~(\ref{deltasca}) with $\theta=1/3$ and $z=1$.
   The dotted lines show fits of the data for the largest chains to
   the polynomial $c_1 \ell^{-1/3} + c_2 \ell^{-2/3}$.  
}
 \label{fig:grad_delta}
\end{figure}

The energy difference of the lowest states is expected to get
suppressed as $O(\ell^{-1/3})$, as predicted by Eq.~(\ref{deltasca})
with $\theta=1/3$ and $z=1$. This is supported by the analysis of the
energy differences $\Delta_\ell = E_1-E_0$ and
$\Delta_{\ell,2}=E_2-E_0$. As shown in Fig.~\ref{fig:grad_delta},
their data are consistent with an asymptotic behavior
\begin{equation}
\Delta_{\ell,\#} \approx c_1 \ell^{-1/3} + c_2 \ell^{-2/3} + ...
\label{deltalda}
\end{equation}

\begin{figure}
\centering
\includegraphics{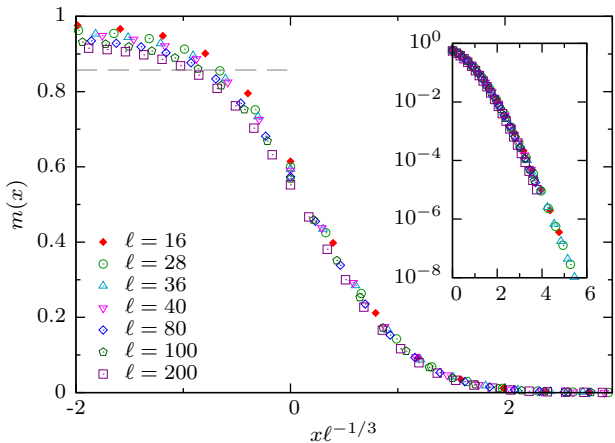}
 \caption{Scaling of the local magnetisation $m(x)$ of the $q=10$
   Potts chain in the presence of a linear magnetic field $h_x$.  The
   dashed line shows the expected left asymptotic value $m_c$,
   corresponding to $f_m(X)=1$ for $X\to -\infty$,
   cf. Eq.~(\ref{magsccl}). This is slowly approached by the data, due
   to the expected $O(\ell^{-1/3})$ corrections.  The convergence
   appears much faster for $x>0$.  The inset shows the data for $x>0$
   in logarithmic scale.}
 \label{fig:grad_mag}
\end{figure}

\begin{figure}
\centering
\includegraphics{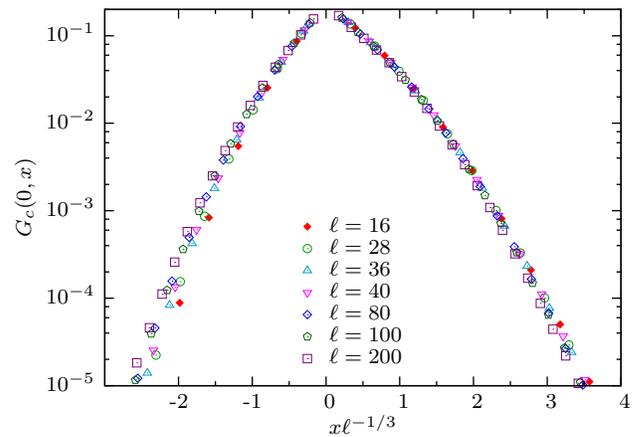}
 \caption{Scaling of the connected two-point function $G_c(0,x)$,
   cf. Eq.~(\ref{gconn}), in the presence of a linear $h_x$.  }
 \label{fig:grad_twop}
\end{figure}

\begin{figure}
\centering
\includegraphics{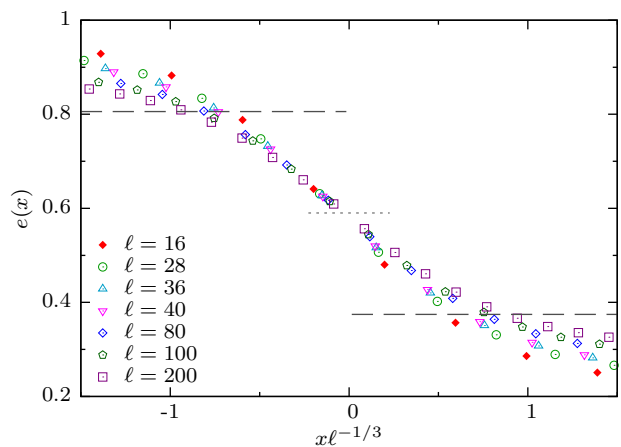}
 \caption{Scaling of the energy density of the $q=10$ Potts chain in
   the presence of a linear $h_x$.  The data appear to approach an
   asymptotic scaling curve supporting the scaling behavior
   (\ref{enesca}).  The dashed lines show the expected asymptotic
   values $e_\pm$ of the scaling function $f_e$,
   cf. Eq.~(\ref{limxfe}).  The central dotted line indicates the
   average value $e_a=(e_+ + e_-)/2$, which seems to be approached by
   the data at $x=0$.  Scaling corrections are clearly observed, in
   particular far from the center; they are consistent with the
   expected (slow) $O(\ell^{-1/3})$ suppression.  }
 \label{fig:grad_ende}
\end{figure}

The data of Figs.~\ref{fig:grad_mag} and \ref{fig:grad_twop}, for the
local magnetization $m(x)$ and the two-point function $G(0,x)$
respectively, appear to approach asymptotic curves when they are
plotted versus $x/\ell^{\theta}$, supporting the scaling behaviors
(\ref{magsccl}) and (\ref{mp1o}).  Scaling corrections are also
clearly observed, which should get asymptotically suppressed by powers
of $\ell^{-\theta}$.  Fig.~\ref{fig:grad_ende} shows the scaling of
the energy density, which support the scaling ansatz derived in
Sec.~\ref{scalFOQT}, given by Eqs.~(\ref{enesca}) and (\ref{limxfe}).

Let us finally note the similarity of these scaling behaviors with
those observed at the first-order classical transition of
two-dimensional Potts models in the presence of a gradient temperature
along one of the spatial directions, with the other one taken to
infinity~\cite{BDV-14}.  Actually, this should not be considered as
unexpected, because the quantum Potts chain and the classical
two-dimensional Potts model are somehow related by a
quantum-to-classical mapping.

\section{Conclusions}
\label{conclusions}

We have shown that scaling phenomena emerge at FOQTs in the presence
of inhomogeneous conditions, such as those arising from a
space-dependent external field, e.g.  $h_x \approx x/\ell$ where
$\ell$ is a length scale.  In particular, we argue that these scaling
phenomena occur in the transition region where the space-dependent
parameter $h(x)$ assumes the value $h_c$ corresponding to the FOQT of
the homogenous system.

We put forward scaling ansatzes to describe the behavior at the
crossover space region where the system effectively changes its phase,
and the typical discontinuities of the FOQT get smoothed out,
i.e. when the system is effectively probing the mixed phase.  This
scaling behavior is characterized by a {\em critical exponent}
$\theta$, cf. Eq.~(\ref{theta}), which tells us how the length scale
$\xi$ of the observables in the crossover region scales with the
length scale $\ell$ of the inhomogeneous field, i.e. $\xi\sim
\ell^\theta$.  The exponent $\theta$ depends on some general features
of the external field giving rise to the inhomogeneity, such as the
effective power law of the space dependence at the transition point
and the way it is coupled to the system variables.  This scaling
behavior is such that the typical singularities of FOQT must be
recovered in the limit $\ell\to\infty$ where the system tend to become
homogenous.  Generally $\theta<1$, approaching one in the limit of an
infinite power law, i.e. $p\to\infty$ in Eq.~(\ref{hip}), where the
inhomogeneous scaling behavior must match the FSS behavior of
homogenous systems with appropriate boundary
conditions~\cite{CNPV-14,CNPV-15}.

We provide numerical evidence of such scaling phenomena for two
classes of FOQTs. We consider the FOQT of quantum Ising chains, which
are driven by a parallel magnetic field when the system is in the
ferromagnetic phase, and those of the $q$-state Potts chain for $q>4$
which is driven by an even temperature-like parameter with a
discontinuity in the ground-state energy density.

Our approach is quite general: the results can be straightforwardly
extended to other systems undergoing FOQTs and other sources of
inhomogeneities smoothing out the singularities of the transition.

These peculiar inhomogeneous scaling phenomena should be observable in
experiments of physical systems, requiring essentially the possibility
of measuring local quantities and controlling/tuning the length-scale
of the inhomogeneity.  Such conditions may be realized in cold-atom
experiments, in particular in optical lattices, when the atomic system
is such to have a FOQT in homogenous conditions, but the space
dependence of the effective chemical potential (arising from the trap)
smooths out its discontinuities.  Around this region we should observe
a crossover region with the scaling features put forward in this
paper.  For example, FOQT lines are expected in the zero-temperature
phase diagrams of atomic systems described by multicomponents
Bose-Hubbard models~\cite{BDZ-08}, with spin-orbit coupling and
synthetic gauge fields, see, e.g.,
Refs.~\onlinecite{Jeckelmann-02,BRS-09,RDSG-12,PMVPFR-14,PCMS-14}.

\end{document}